\documentclass[twocolumn,prb,showpacs]{revtex4}
\usepackage{psfrag}
\usepackage{graphicx}
\usepackage{amsmath}
\usepackage{amssymb}

\begin{document}
\title{Screening effects in a density functional theory based description of molecular junctions in the Coulomb blockade regime}
\author{R. Stadler$^1$, V. Geskin$^2$ and J. Cornil$^2$}
\affiliation{$^{1}$Department of Physical Chemistry, University of Vienna, Sensengasse 8/7, A-1090 Vienna, Austria\\
$^{2}$Laboratory for Chemistry of Novel Materials, University of Mons-Hainaut, Place du Parc 20, B-7000 Mons, Belgium}

\date{\today}

\begin{abstract}
We recently introduced a method based on density functional theory (DFT) and non-equilibrium Green's function techniques (NEGF) for calculating the addition energies of single molecule nano-junctions in the Coulomb blockade (CB) regime. Here we apply this approach to benzene molecules lying parallel and at various distances from two aluminum fcc (111) surfaces, and discuss the distance dependence in our calculations in terms of electrostatic screening effects. The addition energies near the surface are reduced by about a factor of two, which is comparable to previously reported calculations employing a computationally far more demanding quasi-particle description. 
 \end{abstract}
\maketitle

A key issue in the emerging field of molecular electronics is the description of electron transport between nanoscale contacts, for which considerable progress has been recently achieved at the experimental level~\cite{molelect}. Theoretically, two limiting regimes can be distinguished, namely, coherent transport (CT) for strong coupling between the molecule and the electrodes and Coulomb blockade (CB) for weak coupling. The CB regime is best described by stability diagrams, where frontiers between low- and high-conductivity domains in bias and gate voltage coordinates are reflected by diamond-like shapes~\cite{kouwen,kubatkin,nitzan}. For a proper description of these diagrams, the energy difference between the ionization and affinity levels of the inserted quantum dot or single molecule (commonly referred to as addition energies E$_{add}$) has to be evaluated.

The first-principle non-equilibrium Green's function (NEGF) methods~\cite{atk,xue,sanvito,kristian} combined with density functional theory (DFT), which have been successfully used for the CT regime, are not so straightforward to apply for electron transfer under CB conditions, since an integer charge is transferred and results in a relaxation of the electronic structure of the central molecule. In principle only a many-body approach provides a general solution to the latter problem~\cite{greer,datta,ratner}, and even quasiparticle calculations based on the GW approximation~\cite{kristian1} were found to not fully capture the impact of local spin and charge fluctuations in the CB regime~\cite{GW}. The suitability of a standard DFT framework for electron transport in both the CB and CT regimes was also debated~\cite{sanvito1,koentopp}, due to the self-interaction (SI) of electrons~\cite{perdew} in a Kohn-Sham framework (KS) and the lack of a derivative discontinuity (DD)~\cite{perdew1} in the evolution of KS-eigenenergies.
 
As outlined above a main source of discrepancy with a DFT description of weakly coupled nanostructures relies on the fact that the gap between the highest occupied (HOMO) and lowest unoccupied (LUMO) molecular orbital eigenenergies in a single particle KS scheme does not match in general the total energy difference between the ground state and lowest charged states when the size of the HOMO-LUMO gap is finite~\cite{kohn}. This mismatch has been recently addressed in realistic calculations of E$_{add}$ with standard DFT techniques in three different ways: i) For metal particles of finite size, a modified KS gap has been introduced, where the energetic difference between the HOMO (or LUMO) for charged and uncharged clusters has been directly taken into account~\cite{capelle}; ii) For the description of the HOMO-LUMO gap in C$_{60}$-metal interfaces, the charging energy has been obtained by using a constrained DFT formalism~\cite{louie}, where the occupation of hand-picked orbitals can be defined as a constraint in the input~\cite{wu}; iii) Within a NEGF-DFT framework, E$_{add}$ has been defined via threshold values of an external gate voltage V$_{gate}$ determined via a midpoint integration rule from induced charge transfer between small molecules (H$_2$ and benzene) and lithium wires~\cite{first}.

\begin{figure}
    \includegraphics[width=0.95\linewidth,angle=0]{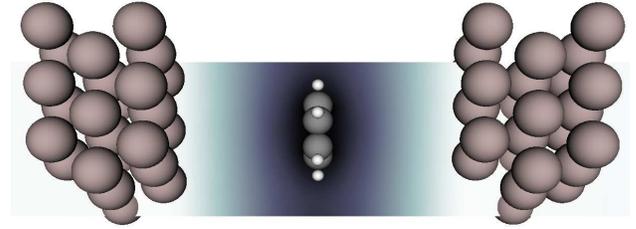}
      \caption[cap.Pt6]{\label{fig.strucpot} (Color online) Geometry and shape of the applied gate potential for a benzene molecule lying parallel and weakly coupled to two Al fcc (111) surfaces for a distance of 8 \AA. The profile of V$_{gate}$(taken from the differences in spatial resolution for calculations at 10V and 0V) is shown with grey shading, with its maximum located in the black regions.} 
    \end{figure}

In this work, we adopt the approach introduced in Ref.~\cite{first} and apply it to calculate E$_{add}$ for a benzene molecule lying parallel and at various distances from two aluminum fcc (111) surfaces (see Fig.~\ref{fig.strucpot}). We focus in particular on the dependence of E$_{add}$ on the distance between the central molecule and the two electrodes, and argue that our method correctly describes screening effects that lead to a reduction of the molecular gap due to the rearrangement of the electronic structure at the Al surfaces. This is supported by two key findings: i) the distance dependence of E$_{add}$ (with a correction term accounting for the geometric capacitance~\cite{capacitance}) is coherent with a purely electrostatic model for screening based on image charges~\cite{kristian2}; ii) The magnitude of screening is comparable to values obtained for a variety of systems with GW~\cite{hybertsen}, constrained DFT~\cite{louie} and a recently developed quantum-chemical approach~\cite{flensberg}, where electrodes have been treated as a classical shape-dependent continuum, allowing to reproduce accurately the experimental data reported in Ref.~\cite{kubatkin}.

Fig.~\ref{fig.strucpot} displays the system on which we performed the NEGF-DFT calculations with the commercially available ATK software~\cite{atk1}. The scattering region contains three layers of a 3x3 unit cell of Al on each side of the benzene molecule and three additionial layers on each side of the left and right electrode regions, respectively, where a 3x3 {\bf k}-point grid has been used for the sampling in the transverse Brillouin plane. All atoms in the Al layers have been left in their truncated bulk positions for the experimental lattice constant of 4.05 \AA . A double-zeta polarized (DZP) and single-zeta (SZ) basis set have been used for the molecule and Al surfaces, respectively, and LDA is chosen for the exchange-correlation (XC) functional. The Keldysh formalism~\cite{atk} allows for a self-consistent solution for the electron density of the open system as a whole for every value of the external gate potential V$_{gate}$~\cite{first}. The shape of the effective potential generated by V$_{gate}$ is illustrated as gray shades in Fig.~\ref{fig.strucpot}.

  \begin{figure}   
  \includegraphics[width=0.95\linewidth,angle=0]{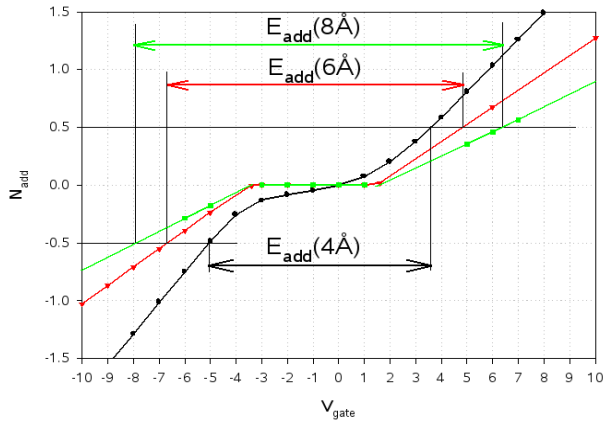}   
  \caption[cap.Pt6]{\label{fig.method}(Color online)Evolution of the added/removed electrons on the benzene molecule N$_{add}$ (obtained by spatial integration of the electron density) as a function of V$_{gate}$ for three distances between the molecule and Al surfaces (d$_{AlB}$=4, 6 and 8 \AA \ ). The values of E$_{add}$ reported elsewhere in this article are taken from such plots. 
}    
  \end{figure}     

In Fig.~\ref{fig.method} we illustrate how we determine the addition energy from NEGF-DFT calculations, E$^{calc}_{add}$, for varying aluminum-benzene distances d$_{AlB}$, 
\begin{eqnarray}
\label{eqn.add}
E^{calc}_{add}&=\int_{0}^1dN_{add}V_{gate}(N_{add})-\int_{-1}^0dN_{add}V_{gate}(N_{add})= \nonumber \\
&V_{gate}(N_{add}=+0.5e)-V_{gate}(N_{add}=-0.5e). \  \ 
\end{eqnarray}

The first line in Eq.~\ref{eqn.add} is general in electrostatics and makes use of the relationship between a total energy E, a voltage V, and a charge N. In the second line, explicit threshold voltages obtained from NEGF-DFT calculations are introduced, and E$^{calc}_{add}$ is evaluated from the gate voltages required to add or subtract half an electron to the molecule. The relevance of N$_{add}$=$\pm$0.5 comes from the integration only; the midpoint rule applies exactly in view of the linearity of the dependence of N$_{add}$ on V$_{gate}$ in DFT~\cite{sanvito1,perdew1,first}, and since we calculate E$^{calc}_{add}$ as the energy corresponding to the integral of V$_{gate}$ over a transferred charge of $\pm$1. This energy represents the input required for the transfer of one electron from the molecule to the two electrodes or vice versa in terms of the external potential V$_{gate}$ inducing this transfer. We stress that this method includes screening effects implicitly, as evidenced by the fact that the results do depend on d$_{AlB}$ (see Fig.~\ref{fig.method}). 

One should keep in mind that E$^{calc}_{add}$ consists of a sum of two terms, the first related to the modified molecular HOMO-LUMO gap in the junction, the second to the electrostatic capacitance of the metallic electrodes~\cite{vanderzant}. The latter is usually referred to as the geometric capacitance contribution to the charging energy in the literature~\cite{capacitance} and will be denoted E$_{geom}$ in the following; it can be safely neglected in the analysis of CB experiments on single molecule junctions since it scales with d$_{AlB}$/A and the area of the electrodes A is usually well above tens of $\mu$m$^2$, whereas the distance d$_{AlB}$ between the molecule and the electrode surfaces is in the \AA \ range. This is not the case, however, in our calculations. Because we apply periodic boundary conditions to the electronic structure in the plane perpendicular to the transport direction, the finite charges are transferred from the molecule to the metal surface in each unit cell. This means that A is defined by only nine atoms in the plane and only d$_{AlB}$ is of similar size as in the experiments, and as a consequence E$_{geom}$ reaches the same order of magnitude than the energetic contribution from the molecular gap. In order to make our results meaningful with respect to experimental values, we define a corrected addition energy as
\begin{eqnarray}
\label{eqn.correct}
E^{corr}_{add} = E^{calc}_{add} - E_{geom} = E^{calc}_{add} - \frac{1}{2} \frac{e^2 (d_{AlB}-x_0)}{2 \epsilon_0 A}
\end{eqnarray}
where e denotes the elementary charge and $\epsilon_0$ the dielectric constant of vacuum. Since the system is equivalent to two capacitors in parallel, the charging energy E$_{geom}$ contains an additional factor $\frac{1}{2}$. Our expression for E$_{geom}$ is rather approximate in the sense that it amounts to replacing the molecule by a third metallic electrode with the same area A as the source and drain electrodes. In Eq.~\ref{eqn.correct} d$_{AlB}$-x$_0$ accounts for the difference in position between the planes of the nuclei and the electrons focal points due to spilling effects, which defines the position of the surface in any purely electrostatic (and therefore not atomistic) model and corresponds to the image plane in the model for screening we introduce below~\cite{kohn1}.

\begin{table}
\caption{\label{Eadd.tab}Corrected addition energies E$^{corr}_{add}$ for three distances d$_{AlB}$ between the Al fcc (111) surfaces and the central benzene molecule; E$_{geom}$ has been subtracted for a meaningful comparison with experiments. The uncorrected values E$^{calc}_{add}$ are also given in parentheses. We provide in two additional columns the corresponding data for systems with Al and Li atomic chains as electrodes in order to connect our discussion to Ref.~\cite{first}. The lattice constants have been chosen as a$_{Al}$ = 2.39 and a$_{Li}$ = 2.9 \AA \ and the areas of the unit cell perpendicular to the wires as A$_{Al}$ = 4a$_{Al}$x4a$_{Al}$ and A$_{Li}$ = 3a$_{Li}$x3a$_{Li}$. The position of the capacitor planes used for determining E$_{geom}$ is taken as x$_0$=2.0 \AA \ for both Al fcc and Al wire electrodes and x$_0$=2.3 \AA \ for the Li wires in accordance with the differences in interlayer spacings~\cite{kohn1}. All values for E$^{corr}_{add}$ and E$^{calc}_{add}$ are given in eV.}
\vspace{0.5 cm}
{\centering
\begin{tabular}{c|c|c|c} \hline
d$_{AlB}$ & Al fcc (111) & Al wire & Li wire \\
\hline
4 \AA \ & 7.27 (8.69) & 8.26 (9.25) & 8.78 (9.80) \\
6 \AA \ & 8.68 (11.51) & 10.18 (12.16) & 9.82 (12.03) \\
8 \AA \ & 10.02 (14.27) & 11.19 (14.16) & 10.99 (14.40) \\
\hline
\end{tabular}\par}\end{table}

We collect in Table~\ref{Eadd.tab} the values for E$^{corr}_{add}$ for three different distances d$_{AlB}$ and contrast them with calculations for the one-dimensional systems studied in Ref.~\cite{first}. For the largest distance of 8 \AA , the results for Li and Al wires as electrodes come rather close to the limiting case (E$^0_{gap}$ = 11.54 eV, as calculated from total energy differences for charged and neutral benzene molecules~\cite{first}), whereas the presence of the surface induces a gap reduction for Al fcc. A decrease in d$_{AlB}$ reduces E$^{corr}_{add}$ for all three types of electrodes due to screening effects which are distance dependent; the HOMO-LUMO gap at 4 \AA \ represents 63 \%, 72  \% and 76 \% of E$^0_{gap}$ with the Al surfaces, Al wires and Li wires, respectively. These numbers are comparable to those found for other systems~\cite{louie,hybertsen,flensberg}. Although screening is usually associated with the interaction of charges with surfaces, a similar albeit quantitatively smaller effect can also be observed in Table~\ref{Eadd.tab} for the wire electrodes.

In order to validate the E$^{corr}_{add}$ (d$_{AlB}$) values provided by our approach, we have also estimated the corresponding numbers from an image charge model. For that purpose we define the molecular contribution to E$_{add}$ (the capacitative term E$_{geom}$ does not enter the picture here, since it is a correction for the finiteness of the unit cell, whereas the image charge model assumes an infinite surface by definition) as

\begin{eqnarray}
\label{eqn.screen}
E^{image}_{add}=(E(N+1)-E(N))-(E(N)-E(N-1))= \nonumber \\
E^0_{gap} + \Sigma(N+1)+\Sigma(N-1) = E^0_{gap} - E_{screen} \  \
\end{eqnarray}

with E(N$\pm$1)=E$^0$(N$\pm$1)+$\Sigma$(N$\pm$1). E$^0_{gap}$ denotes the difference between the electron affinity and ionization potential of the free molecule and $\Sigma$ the correction due to screening (i.e. the image energies associated to the charges that are calculated following the detailed recipe given in the supplementary information of Ref.~\cite{kristian2}). $\Sigma(N)$=0, since there are no screening effects when the benzene molecule is neutral, because it does not exhibit any polar bonds. 

\begin{figure}
  \includegraphics[width=0.95\linewidth,angle=0]{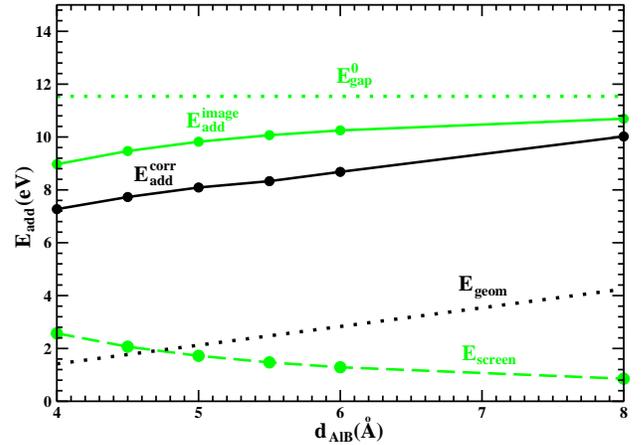}
  \caption[cap.Pt6]{\label{fig.results}(Color online) Evolution of E$^{corr}_{add}$ as a function of d$_{AlB}$, as extracted from NEGF-DFT calculations and compared to E$^{image}_{add}$ (for its definition see Eq.~\ref{eqn.screen}). E$^0_{gap}$ = 11.54 eV as calculated in Ref.~\cite{first}.
}
  \end{figure}

We compare in Fig.~\ref{fig.results} the distance dependence of E$_{add}$ as obtained from NEGF-DFT calculations via Eqs.~\ref{eqn.add} and~\ref{eqn.correct} to the results provided by the image charge model, i.e. we test the assumption
 
\begin{eqnarray}
\label{eqn.sum}
E^{image}_{add} (d_{AlB}) \approx E^{corr}_{add} (d_{AlB}).
\end{eqnarray}

The deviations between E$^{image}_{add}$ and E$^{corr}_{add}$ are less than 20 percent of the total value of E$^{corr}_{add}$ for the range of d$_{AlB}$ values under consideration. This is quite remarkable given the approximative nature of: i) E$_{screen}$ (defining the d$_{AlB}$ dependence of E$^{image}_{add}$ via Eq.~\ref{eqn.screen}) since only the charge distribution inside the molecule has been described realistically by using Mulliken charges~\cite{kristian2}; and ii) E$_{geom}$ (contributing to E$^{corr}_{add}$ via Eq.~\ref{eqn.correct}) for which the molecule and metal surfaces have been replaced by capacitor planes without taking into account any details of their atomic structures. When d$_{AlB}$ gets smaller (close to 4 \AA ), it has also to be considered that wavefunction overlap, which is not included in electrostatic models, starts to play a role so that the agreement between E$^{image}_{add}$ and E$^{corr}_{add}$ is expected to be better at large distances.

Finally, we want to position our work in the context of the other recently proposed methods for the theoretical description of CB experiments with single molecule junctions. The modified KS scheme of Ref.~\cite{capelle} is based on finite systems and therefore not directly suitable to study screening effects. Our method differs from those in Refs.~\cite{hybertsen} and~\cite{flensberg} by its level of accuracy; although inferior to a full quasi-particle description~\cite{hybertsen}, which can treat only rather small systems, our approach is preferable to a semi-empirical method~\cite{flensberg}, where the predictive power is limited by the need to find suitable parameters. The technique in Ref.~\cite{louie} based on constrained DFT~\cite{wu} is rather close to our approach in the sense that it also enforces the occupation of molecular orbitals and calculates E$_{add}$ from the energy required to uphold this charging. However, while we apply an external gate voltage V$_{gate}$, let the electron density relax as a function of it and determine E$_{add}$ from threshold values for V$_{gate}$, the occupation of the HOMO/LUMO are fixed manually in Ref.~\cite{louie}, thus introducing a certain amount of arbitrariness (at least for not so weakly coupled systems), and the key quantity is the gradient of orbital eigenenergy with its occupation. We stress that all these methods find a gap reduction due to screening effects by about a factor of two, in good agreement with our results.

In summary, we have extended a recently introduced method for the calculation of addition energies E$_{add}$ for single-molecule junctions in the CB regime~\cite{first} by describing in a more realistic way the electrode surfaces. This paves the way towards NEGF-DFT based predictions of E$_{add}$ for junctions characterized in recent experimental studies~\cite{kubatkin}, where screening effects are likely to play a major role. We analyzed the distance dependence of E$_{add}$ in comparison to an image charge model and to other techniques developed for determining E$_{add}$ and found an overall good agreement between all approaches suggesting a reduction of the electronic gap by up to 50 \% in molecular junctions.

This research has been supported by the European projects SINGLE (FP7/2007-2013, no. 213609) and MODECOM (NMP3-CT-2006-016434), the Interuniversity Attraction Pole Program of the Belgian Federal Science Policy Office (PAI 6/27), and the Belgian National Fund for Scientific Research (FNRS). We are especially indebted to Dr. J. A. Torres for his advice. J.C. is a Research Associate of FNRS. R.S. is currently supported by the Austrian Science Fund FWF, project Nr. P20267.


\bibliographystyle{apsrev}

\end{document}